\documentclass[AMS,STIX1COL]{WileyNJD-v2}
\usepackage{moreverb}
\usepackage{mathtools}
\usepackage{bm}
\usepackage{comment}
\usepackage{placeins} %used to prevent latex from mixing up bibliography and figures

\newcommand\BibTeX{{\rmfamily B\kern-.05em \textsc{i\kern-.025em b}\kern-.08em
T\kern-.1667em\lower.7ex\hbox{E}\kern-.125emX}}

\articletype{Original Article}%

\begin{document}

\title{A data-driven approach for 2D vorticity PDF equations by a new conditional average estimation}

\author[1]{Qian Huang*}

\author[2,3]{Simon Görtz}

\author[2,3]{Paul Hollmann}

\author[2,3]{Johannes Conrad}

\author[1]{Christian Rohde}

\author[2,3]{Martin Oberlack}

\authormark{Huang, Görtz, Hollmann \textsc{et al}}

\address[1]{\orgdiv{Institute of Applied Analysis and Numerical Simulation}, \orgname{Stuttgart University}, \orgaddress{Pfaffenwaldring 57,
70569 Stuttgart}, \country{Germany}}

\address[2]{\orgdiv{Chair of Fluid Dynamics}, \orgname{Technical University of Darmstadt}, \orgaddress{Otto-Berndt-Stra{\ss}e 2, 64287 Darmstadt, \country{Germany}}}

\address[3]{\orgdiv{Centre for Computational Engineering}, \orgname{TU Darmstadt}, \orgaddress{Dolivostra{\ss}e 15, 64293 Darmstadt, \country{Germany}}}

\corres{*Qian Huang, \email{Qian.Huang@ mathematik.uni-stuttgart.de}}

%\presentaddress{Present address}

\abstract[Abstract]{We consider the statistics 
for the vorticity field in two-dimensional homogeneous isotropic turbulence (HIT).
First, we exploit  the invariance properties 
to derive dimensionally reduced governing equations for the one-point and two-point probability density functions (PDFs). These take the form of   linear kinetic transport equations, but with an unclosed operator in terms of a conditional average.\\
To solve the PDF equation numerically we suggest  a hybrid data-driven method that
relies on  carefully selected samples of DNS data and a  sampling estimator for the 
conditional  average. The  method is applied to DNS data for both decaying and forced HIT, demonstrating good agreement  with the direct evaluation of the PDFs using the DNS data.}


\maketitle

\section{Introduction}\label{sec1}
%\todo{Engineer's perspective :) :}\\
The statistical description of turbulence remains challenging due to the nonlinear and multiscale nature of the governing equations. A formally exact framework is provided by the Lundgren-Monin-Novikov (LMN) hierarchy \citep{Lundgren1967,MoninYaglom1975,Novikov1965}, which yields evolution equations for multi-point probability density functions (PDFs) of the flow variables. Equivalent to the hierarchy of moment equations, the LMN approach shifts the closure problem to conditional averages of dynamical quantities, such as pressure gradients and dissipation, that depend on higher-order statistics. In practice, this formulation is often recast in terms of single-point PDFs, where the unclosed terms appear as conditional expectations that may be modelled or extracted from direct numerical simulations \citep{FriedrichPeinke1997,Pope2000}.

The PDF framework is particularly well suited to vorticity-based descriptions. In two-dimensional incompressible flow, the dynamics reduces to the scalar vorticity equation, in which the absence of vortex stretching leads to a qualitatively different statistical structure compared to three-dimensional turbulence. Specifically, vorticity is materially advected and diffused, while remaining nonlinearly coupled to the velocity field through the Biot-Savart law. This simplification makes two-dimensional homogeneous isotropic turbulence a natural setting for investigating PDF hierarchies and closure strategies.

Two-dimensional turbulence is characterised by a dual cascade, with an inverse transfer of energy to large scales and a direct cascade of enstrophy to small scales \citep{Batchelor1969,Kraichnan1967}, accompanied by the formation of long-lived coherent vortices \citep{McWilliams1984,Weiss1991}. These structures strongly influence vorticity statistics, leading to pronounced non-Gaussian behaviour and intermittency \citep{FalkovichLebedev1994,ParetTabeling1998}. Within the LMN framework, such effects are encoded in conditional averages describing vorticity transport and dissipation \citep{FRIEDRICH2012929,WilczekFriedrich2009}. Recent work has demonstrated that these quantities can be quantified from numerical simulations, enabling a data-driven assessment of closure assumptions and providing insight into the interplay between coherent structures and turbulent statistics. For example, Wilczek \& Friedrich (2009) evaluate conditional averages of stretching and diffusion in the single‑point vorticity PDF equation using direct numerical simulations, linking dynamical effects to the emergence of non‑Gaussian statistics \citep{WilczekFriedrich2009}. Similarly, Friedrich et al. (2012) derive two‑point vorticity PDF evolution equations in the LMN framework and use DNS data to study and compare the unclosed terms against Gaussian approximations in the inverse energy cascade of 2D turbulence, interpreting the results in terms of effective vortex interactions \citep{FRIEDRICH2012929,FriedrichVosskuhleKampsWilczek2012}.

Despite these advances, a predictive numerical closure of the PDF hierarchy remains an open problem.
In this paper we follow an intermediate strategy generalizing our previous 
work \cite{huang2025numerical}  which starts from  scalar 
viscous balance laws but not from the incompressible Navier-Stokes equations. 
This hybrid method relies on the direct numerical solution  of the PDF equation 
and approximates the unclosed 
terms by using sampled quantities of
the underlying  system of evolution equations. Precisely, we suggest to use as a 
kernel based estimator the Nadaraya-Watson-type estimator \cite{Bierens_1994} to approximate ther 
the conditional expectation. Our numerical experiments indicate that 
this approach  is robust and  enhances the sample efficiency as compared to constructing the distribution directly from samples, as in Monte Carlo-type methods.
The hybrid method is not restricted to 
one-point  probability
density functions    but can also be 
applied to  two-point probability
density functions. It offers then  a structure-based  way  for capturing correlations as an essential feature in turbulence. 

The rest of the paper is organized as follows. Section~\ref{sec:DNS} describes the direct numerical simulations used for data generation. 
In Section~\ref{sec:pdf}, the vorticity PDF hierarchy is derived under the assumptions of homogeneity and 
isotropy, leading to the one-point transport equations. Section~\ref{sec:num} presents the numerical method, and numerical results are discussed in Section~\ref{sec:result} for both decaying and forced turbulence. Finally, conclusions are drawn in Section~\ref{sec:conclusion}.

%---------------------------------------
\section{Data generation} \label{sec:DNS}
The direct numerical simulations (DNS) performed in this work are based on the open-source pseudo-spectral solver by Lauber \cite{lauber_2d_turbulence_github}, released under the MIT license. The original code was adapted and extended to include the forcing and large-scale damping mechanisms described in this section, as well as additional diagnostics for the present study.

We consider the incompressible Navier-Sokes equations in two dimensions
\begin{equation} \label{eq:navier-stokes}
    \frac{\partial \boldsymbol{u}}{\partial t}
+ (\boldsymbol{u} \cdot \nabla)\boldsymbol{u}
= -\nabla p + \frac{1}{\mathrm{Re}} \Delta \boldsymbol{u},
\end{equation} 
together with the incompressibility constraint
\begin{equation} \label{eq:incompressible}
\nabla \cdot \boldsymbol{u} = 0,
\end{equation}
where $\boldsymbol{u}(\boldsymbol{x},t)$ is the velocity field, $p(\boldsymbol{x},t)$ the pressure. Taking the curl of equations \eqref{eq:navier-stokes} and \eqref{eq:incompressible} yields the 2D forced vorticity transport equation
 \begin{equation}
    \label{eq:forced_vorticity_transport}
    \frac{\partial \omega}{\partial t}+(\boldsymbol u\cdot\nabla)\omega=\frac{1}{\mathrm{Re}}\Delta \omega+F(\boldsymbol{x},t),
 \end{equation}
where $\omega(\boldsymbol{x},t)=\mathrm{curl}\, \boldsymbol{u}(\boldsymbol{x},t) = \partial_x v - \partial_y u \in \mathbb R$ denotes the vorticity field and $F(\boldsymbol{x},t)$ is an added forcing and damping term.

The computational domain is chosen as a doubly periodic square domain $D = [0,2\pi]^2$, ensuring compatibility with the spectral discretization and allowing for an efficient FFT-based implementation. Different simulation setups are considered, varying resolution, Reynolds number, and forcing. An overview of all configurations is given in Table~\ref{tab:dns-params}.
The choice of simulation parameters is guided by resolution requirements and the targeted flow regimes. The spatial resolutions are selected such that the dissipative scales remain well separated from the grid cutoff (i.e.\ $k_{\max}\eta \gtrsim 1$). For the higher Reynolds number cases, the resolution is increased accordingly to account for the extended range of active scales.
For the forced simulations, energy is injected at relatively high wavenumbers in the range $k_{f,\min} \leq |\boldsymbol{k}| \leq k_{f,\max}$, corresponding to small spatial scales. This promotes the development of an inverse energy cascade towards larger scales. To counteract the accumulation of energy at the largest scales, a linear damping is applied in the low-wavenumber range $k_{d,\min} \leq |\boldsymbol{k}| \leq k_{d,\max}$, mimicking large-scale friction.
The damping coefficient $\beta = 0.15$ is chosen to remove energy at large scales while remaining weak compared to nonlinear transfer processes. The relaxation parameter $\tau_{\mathrm{relax}} = 0.6$ sets the timescale over which the total kinetic energy is controlled and is selected to ensure statistical stationarity without suppressing turbulent fluctuations.

To solve equation \eqref{eq:forced_vorticity_transport}, a pseudo-spectral method is employed. On the periodic domain $D=[0,2\pi]^2$, the vorticity field $\omega(\boldsymbol{x},t)$ is represented by a discrete Fourier series
\begin{equation}
    \hat{\omega}(\boldsymbol{k},t) = \frac{1}{|D|} \int_{D} \omega(\boldsymbol{x},t)\, e^{-i \boldsymbol{k}\cdot\boldsymbol{x}} \, \mathrm d \boldsymbol{x}, 
    \quad
    \omega(\boldsymbol{x},t) = \sum_{\boldsymbol{k}\in\mathbb{Z}^2} \hat{\omega}(\boldsymbol{k},t)\, e^{i \boldsymbol{k}\cdot\boldsymbol{x}},
\end{equation}
where $\boldsymbol{k}=(k_x,k_y)\in\mathbb{Z}^2$ denotes the discrete wavevector and $|D|=(2\pi)^2$ is the domain size. In practice, the Fourier transforms are evaluated using fast Fourier transforms (FFT). In the following, quantities in Fourier space are indicated by a hat.

Applying the Fourier transform to \eqref{eq:forced_vorticity_transport} yields the vorticity equation 
\begin{equation}
\frac{\partial \hat{\omega}}{\partial t}
+ \widehat{(\boldsymbol{u}\cdot\nabla \omega)}
= -\frac{1}{\mathrm{Re}} |\boldsymbol{k}|^2 \hat{\omega}
+ \hat{F}(\boldsymbol{k},t)
\end{equation}
where $\hat{\omega}(\boldsymbol{k},t)$ denotes the Fourier coefficients of the vorticity field.
In this formulation, spatial derivatives are evaluated as algebraic multiplications by $i\boldsymbol{k}$ in Fourier space, while the viscous term becomes diagonal, since the Laplacian reduces to a scalar multiplication $-|\boldsymbol{k}|^2$ acting independently on each Fourier mode. 
The nonlinear advection term $\widehat{(\boldsymbol{u}\cdot\nabla \omega)}$ corresponds to a convolution in Fourier space and is therefore computed in physical space using the pseudo-spectral approach: the velocity field is obtained from the streamfunction (see \eqref{eq:streamfunction}), the nonlinear term is evaluated pointwise, and the result is transformed back to Fourier space. To avoid an accumulation of aliasing errors, the $2/3$-dealiasing rule is applied.
The vorticity field is advanced in time in Fourier space in terms of its spectral coefficients $\hat{\omega}(\boldsymbol{k},t)$ using a third-order total variation diminishing (TVD) Runge-Kutta method \cite{SHU1988}. The time step $\Delta t$ is dynamically adapted based on a Courant–Friedrichs–Lewy (CFL) condition, accounting for both advective and viscous stability constraints. The velocity field is recovered via the stream function using
\begin{equation} 
    \label{eq:streamfunction}
    \hat{\psi}(\boldsymbol{k},t) = -|\boldsymbol{k}|^{-2} \hat{\omega}(\boldsymbol{k},t),
    \quad \text{with} \quad 
    \boldsymbol{u} = \left( \frac{\partial \psi}{\partial y}, \, -\frac{\partial \psi}{\partial x} \right).
\end{equation}

The forcing is defined in Fourier space, where it acts selectively on prescribed wavenumber bands. Specifically, we consider a linear forcing \cite{OVERHOLT1998} and damping of the form
\begin{equation}
    \hat{F}(\boldsymbol{k},t) =
    \begin{cases}
        \alpha(t)\, \hat{\omega}(\boldsymbol{k},t), & k_{f,\min} \leq |\boldsymbol{k}| \leq k_{f,\max}, \\
        -\beta\, \hat{\omega}(\boldsymbol{k},t), & k_{d,\min} < |\boldsymbol{k}| \leq k_{d,\max}, \\
        0, & \text{otherwise},
    \end{cases}
\end{equation}
where the first term injects energy at small scales and the second term introduces linear damping at large scales.
The corresponding forcing in physical space is obtained via inverse Fourier series,
\begin{equation}\label{eq:forcing}
    F(\boldsymbol{x},t)
    = \alpha(t) \sum_{k_{f,\min} \leq |\boldsymbol{k}| \leq k_{f,\max}} 
    \hat{\omega}(\boldsymbol{k},t)\, e^{i \boldsymbol{k}\cdot\boldsymbol{x}} 
    - \beta \sum_{k_{d,\min} < |\boldsymbol{k}| \leq k_{d,\max}} 
    \hat{\omega}(\boldsymbol{k},t)\, e^{i \boldsymbol{k}\cdot\boldsymbol{x}}.
\end{equation}
which yields the integral representation over the respective forcing and damping bands. This spectrally localized forcing combined with large-scale damping is a common approach in two-dimensional turbulence simulations (see e.g. \cite{XIAO_WAN_CHEN_EYINK_2009}), while linear forcing schemes proportional to the flow field provide a mechanism to maintain statistical stationarity \cite{OVERHOLT1998}.

In the present work, the damping coefficient is fixed at $\beta = 0.15$.
The forcing amplitude $\alpha(t)$ is dynamically determined at each time step of the energy budget to maintain a statistically stationary state. Specifically, it is chosen such that the energy input balances viscous dissipation while relaxing the total kinetic energy towards its initial value,
\begin{equation}
    \left. \frac{\mathrm d E}{\mathrm d t} \right|_{\mathrm{forcing}}
    = \alpha(t)\, 2E_f(t)
    = \left(E(0) - E(t)\right) /\tau_\mathrm{relax} + \varepsilon_\nu.
\end{equation}
The total kinetic energy $E(t)$ is defined by
\begin{equation}
    E(t) = \frac{1}{2} \int_D \left( u^2 + v^2 \right) \, \mathrm d \boldsymbol{x}
    = \frac{1}{2} \sum_{\boldsymbol{k}} |\boldsymbol{k}|^{-2} |\hat{\omega}(\boldsymbol{k},t)|^2,
\end{equation}
and $\varepsilon_\nu$ is the viscous dissipation rate,
\begin{equation}
    \varepsilon_\nu(t) = \frac{1}{\mathrm{Re}} \int_D \omega^2 \, \mathrm d \boldsymbol{x}
    = \frac{1}{\mathrm{Re}} \sum_{\boldsymbol{k}} |\hat{\omega}(\boldsymbol{k},t)|^2.
\end{equation}
The relaxation parameter $\tau_\mathrm{relax}$ controls the rate at which the total kinetic energy is driven towards its initial value $E(0)$. Therefore, it introduces a feedback mechanism that prevents a long-term drift of the energy level. Small values of $\tau_\mathrm{relax} \ll 1$ correspond to strong control, enforcing rapid relaxation towards the target energy, while larger values allow for more pronounced temporal fluctuations and a weaker coupling between forcing and the instantaneous energy state. In the present simulations, $\tau_\mathrm{relax}$ is chosen such that a statistically stationary regime is reached while still permitting natural turbulent variability.
The quantity $E_f(t)$ denotes the kinetic energy contained in the forced wavenumber band,
\begin{equation}
    E_f(t) = \frac{1}{2} \sum_{k_{f,\min} \leq |\boldsymbol{k}| \leq k_{f,\max}} 
    |\boldsymbol{k}|^{-2} |\hat{\omega}(\boldsymbol{k},t)|^2.
\end{equation}
It determines the efficiency with which the linear forcing injects energy into the system, such that the total energy input rate is proportional to $\alpha(t)\,2E_f(t)$.
For reasons of simplicity, we will in the following assume $\alpha(t)$ as externally given. Especially in the following statistical treatment of the problem, $\alpha(t)$ being the link between $\alpha$ and $\omega$ via the energy budget will be omitted, since it leads to non-localities which are difficult to treat in the calculation of conditional averages. 

The initial condition for the vorticity field is constructed following the approach of McWilliams \cite{McWilliams1984}. The streamfunction coefficients are defined as
\begin{equation}
    \hat{\psi}(\boldsymbol{k},0) = c(\boldsymbol{k}) \left(\xi_1 + i\xi_2 \right), \quad \text{with} \quad c(\boldsymbol{k}) =\left(|\boldsymbol{k}| \left[1+\frac{|\boldsymbol{k}|^2}{36}^2\right] \right)^{-1/2},
\end{equation}
where $\xi_1$ and $\xi_2$ are independent Gaussian random variables. A high-wavenumber filter is applied to suppress small-scale contributions, and the field is subsequently normalized to yield a prescribed initial kinetic energy. The initial vorticity field is then obtained via \eqref{eq:streamfunction}.

\begin{table}[h]
\centering
\caption{Simulation parameters for the different HIT DNS cases.\label{tab:dns-params}}
\begin{tabular}{lcccccc}
\hline
\textbf{Sim} & $N_x \times N_y$ & $\mathrm{Re}$ & $\beta$ & $\big[k_{d,\min},k_{d,\max}\big]$ & $\big[ k_{f,\min},k_{f,\max}\big]$ & $\tau_{\mathrm{relax}}$ \\
\hline
Case 1 & $256^2$  & 200 & - & - & - & - \\
Case 2 & $384^2$  & 360 & - & - & - & - \\
Case 3 & $512^2$ & 200 & 0.15 & [0,2] & [15,40] & 0.6 \\
Case 4 & $512^2$ & 360 & 0.15 & [0,2] & [15,40] & 0.6 \\
\hline
\end{tabular}
\end{table}

%---------------------------------------------

\section{Vorticity PDF hierarchy in relative coordinates} \label{sec:pdf}
We consider 2D homogeneous isotropic turbulence in the framework of the vorticity formulation. For this we repeat the 2D vorticity equation, which is given in nondimensionalized form by
\begin{equation}\label{eq:vorticity_eq}
    \frac{\partial \omega}{\partial t}+(\boldsymbol u\cdot\nabla)\,\omega=\frac{1}{\mathrm{Re}}\Delta\omega,
\end{equation}
where vortex stretching is naturally turned off for the 2D case.
We consider the field $\omega(\boldsymbol x,t)$ as statistical and therefore use multi-point vorticity PDFs $f_n$ for its description. By using an arbitrary test function $Q$, the mean value is defined by
\begin{equation}\label{eq:weak_def_pdf}
    \langle Q(\omega(\boldsymbol x_1,t),\ldots,\omega(\boldsymbol x_n,t))\rangle=\int_{\mathbb R^n} Q(\Omega_1,\ldots,\Omega_n)f_n(\Omega_1,\boldsymbol x_1,\ldots,\Omega_n,\boldsymbol x_n,t)\,\mathrm d\Omega_1\mathrm d\Omega_n,
\end{equation}
where $\Omega_k$ refers to the $k$-th sample space vorticity coordinate and the integral is carried out over the full space of sample space velocities. In the following, integral boundaries are omitted for the sake of brevity. Unless otherwise stated, integrations are performed over the entire space of samples.
We further define the conditional PDF including an additional statistical field, here denoted by $p(\boldsymbol x,t)$ with corresponding sample space variable $q$, exemplarily for the one-point case by
\begin{equation}\label{eq:cond_pdf}
    f_{p| \omega}(q|\Omega_1,\boldsymbol x,t)=\frac{f_1(q,\Omega_1,\boldsymbol x,t)}{f_1(\Omega_1,\boldsymbol x_1,t)}
\end{equation}
and the conditional average as
\begin{equation}\label{eq:cond_average}
    \left\langle p(\boldsymbol x,t)|\Omega_1=\omega(\boldsymbol x,t)\right\rangle = \int q \,f_{p| \omega}(q|\Omega_1,\boldsymbol x,t)\,\mathrm d q.
\end{equation}
In the following, we will use the abbreviation $\left\langle p(\boldsymbol x,t)|\Omega_1,\boldsymbol x,t\right\rangle$ for the conditional average.
Using the definition \eqref{eq:weak_def_pdf}, a hierarchy of transport equations for the multi point PDFs can be derived from \eqref{eq:vorticity_eq} as given in \cite{Novikov}. To study HIT, these equations are written in spatial increment coordinates according to $\boldsymbol x=\boldsymbol x_1\,, \boldsymbol r_2=\boldsymbol x_2-\boldsymbol x_1,\ldots \boldsymbol r_n=\boldsymbol x_n-\boldsymbol x_1$. With that, the derivatives become
    \begin{align}\label{eq:spatial_increments}
        \frac{\partial}{\partial \boldsymbol x_1}=\frac{\partial}{\partial \boldsymbol x}-\sum_{k=2}^n\frac{\partial}{\partial \boldsymbol r_n}, \qquad \qquad
        \frac{\partial}{\partial \boldsymbol x_k}=\frac{\partial}{\partial \boldsymbol r_k} \quad \forall \quad k\geq 2.
    \end{align}
Starting from the form given in \cite{Novikov}, we obtain the infinite multi-point PDF hierarchy in relative coordinates as
\begin{align}
    \frac{\partial f_n}{\partial t}=&\frac{1}{4\pi}\left(\nabla_{\boldsymbol x}\cdot\!\!\int\frac{\boldsymbol r_{n+1}}{r_{n+1}^3}\times\int\Omega_{n+1}\boldsymbol{e}_zf_{n+1},\mathrm d\Omega_{n+1}\mathrm d^2\boldsymbol r_{n+1}+\sum_{s=2}^{n}\nabla_{\boldsymbol r_s}\cdot\!\!\int\left(\frac{\boldsymbol r_s-\boldsymbol r_{n+1}}{|\boldsymbol r_s-\boldsymbol r_{n+1}|^3}-\frac{\boldsymbol r_{n+1}}{r_{n+1}^3}\right)\times\int\Omega_{n+1}\boldsymbol{e}_zf_{n+1},\mathrm d\Omega_{n+1}\mathrm d^2\boldsymbol r_{n+1}\right) \nonumber\\
    &-\frac{1}{\mathrm{Re}}\left(\frac{\partial}{\partial \Omega_1}\lim_{\boldsymbol r_3\to 0}+\sum_{s=2}^{n}\frac{\partial}{\partial \Omega_s}\lim_{\boldsymbol r_{n+1}\to \boldsymbol r_s}\right)\Delta_{\boldsymbol r_{n+1}}\int\Omega_{n+1}f_{n+1}\,\mathrm d\Omega_{n+1}. \label{vorticity_hierarchy_rel_coord}
\end{align}
Here, the terms on the RHS in the first line follow from the convective term with the velocity expressed by vorticity using Biot Savart's law. The terms in the second line stem from viscous stresses. Note that these unclosed terms can be rewritten in terms of conditional averages, extending the definitions given by equations \eqref{eq:cond_pdf} and \eqref{eq:cond_average} to the multi-point case. This will be detailed below for the respective one- and two-point equations.
\subsection{Homogeneous Isotropic PDFs}
We consider the field $\omega(\boldsymbol x,t)$ as statistically homogeneous and isotropic, meaning that the averaged quantities and accordingly the PDFs are homogeneous and isotropic, i.e. invariant under translations and rotations of the coordinate system. Accordingly, the PDF, as a scalar density, has the reduced dependency
\begin{subequations}\label{eq:reduced_HIT_Pdf_dependencies}
    \begin{align}
    f_1=f_1(\Omega_1,t),\\
    f_2=f_2(r_2,\Omega_1,\Omega_2,t),\\
    f_3=f_3(r_2,r_3,\phi_{23},\Omega_1,\Omega_2,\Omega_3,t),
\end{align}
\end{subequations}
with $r_k=|\boldsymbol r_k|$ and the relative angle $\phi_{23}=\angle(\boldsymbol r_2,\boldsymbol r_3)$. In continuation, higher-order PDFs introduce two additional spatial coordinates and one sample-space variable per point, i.e., per order. Hence, the dimensional reduction by one rotational and two translational degrees of freedom first becomes apparent at the two-point level and continues for higher PDFs, with the reduction by three dimensions remaining constant at each level.
\subsection{One-point hierarchy}
We begin with the reduction of the one-point hierarchy. 
Introducing homogeneity, i.e. independency of the PDFs from the base point by $\partial /\partial \boldsymbol x=0$, the one-point equation in Cartesian index notation, following from \eqref{vorticity_hierarchy_rel_coord} reduces to 
\begin{equation}\label{eq:1-point_hierarchy-homogeneoues}
    \frac{\partial f_1}{\partial t}=\frac{\nabla_{\boldsymbol r_2}}{4\pi}\cdot\int\frac{\boldsymbol r_2^\perp}{r_2^3}\int\Omega_2 f_2\,\mathrm d\Omega_2\mathrm d^2\boldsymbol r_2 - \frac{1}{\mathrm{Re}} \frac{\partial}{\partial \Omega_1}\lim_{\boldsymbol r_2\to 0} \Delta_{\boldsymbol r_2}\int \Omega_2 f_2 \,\mathrm d\Omega_2,
\end{equation}
with the abbreviation $\boldsymbol r_2^\perp=\boldsymbol r_{2,y}\boldsymbol e_x-\boldsymbol r_{2,x}\boldsymbol e_y$. It is further observed that the convective term completely vanishes since the integral does not depend on $\boldsymbol r_2$ after integrating over the full $\mathbb R^2$ as unbounded domain. Note that the unbounded domain is modelled in the simulation using periodic boundary conditions as discussed in section \ref{sec:DNS}. If these periodic boundary conditions were strictly accounted for in the theory, boundary integral terms would appear in the corresponding Biot-Savart integral. For the sake of simplicity, these are not considered here, instead, an unbounded domain is assumed. Consequently, the non-stationary term is balanced by the viscous term solely.
At this point, we additionally introduce isotropy, meaning that the PDF remains invariant under rotations, with the dependencies as introduced in equations \eqref{eq:reduced_HIT_Pdf_dependencies}. In order to reflect isotropy in the equations, we introduce polar coordinates $(r_k=|\boldsymbol r_k|,\theta_k)$ describing the space increment vector by $\boldsymbol r_k=r_k\,\boldsymbol e_r(\theta_k)$. 
Further, the Laplacian occurring in the viscous term in equation \eqref{eq:1-point_hierarchy-homogeneoues} can simply be replaced by the usual Laplacian in polar coordinates since the basis vectors for the description of the vorticity sample space are unaffected by the change of the coordinate system . We obtain
\begin{equation}\label{eq:reduced_one_point_unclosed}
    \frac{\partial f_1(\Omega_1,t)}{\partial t}=- \frac{1}{\mathrm{Re}} \frac{\partial}{\partial \Omega_1}\lim_{\boldsymbol r\to 0} \int \Omega_2 \left(\frac{\partial^2}{\partial r^2}+\frac{1}{r}\frac{\partial}{\partial r}\right)f_2 \,\mathrm d\Omega_2.
\end{equation}
Note that this equation may formally be closed by the reduction property, i.e. $f_1=\int f_2\mathrm d\Omega_2$. Expressed in the usual notation with conditional averages, \eqref{eq:reduced_one_point_unclosed} is equivalent to
\begin{equation} \label{eq:1pointPDF_HIT}
    \frac{\partial f_1(\Omega_1,t)}{\partial t}=- \frac{1}{\mathrm{Re}} \frac{\partial}{\partial \Omega_1}\left(\langle\Delta \omega(\boldsymbol x,t)|\Omega_1, t\rangle f_1\right).
\end{equation}
Due to the statistical homogeneity of the considered field, it should be pointed out that the conditional average has a reduced dependency.

\subsection{Forced isotropic turbulence}
In the absence of external forcing, energy in homogeneous isotropic turbulence (HIT) is continuously dissipated, leading to a decay of all solutions at large times. This behaviour becomes evident from \eqref{eq:reduced_one_point_unclosed}, where the non-stationary term is solely governed by viscous dissipation.
To sustain a statistically stationary state of HIT, external body forces are commonly introduced in numerical simulations of two-dimensional turbulence (see, e.g., \cite{BoffettaEcke2012, AlexakisBiferale2018}). These forcing terms extend the vorticity transport equation and, consequently, also modify the associated PDF transport equation, as for example shown by Friedrich et al. \cite{FRIEDRICH2012929}. In the following, we demonstrate how the additional unclosed terms arising in the PDF transport equation due to forcing can be approximated using the method presented above. For brevity, we restrict the discussion to the one-point case. With the deterministic forcing term as given by \eqref{eq:forcing}, the one-point PDF evolution equation becomes
\begin{equation} \label{eq:1pointPDF_HIT_forced}
    \frac{\partial f_1(\Omega_1,t)}{\partial t}=- \frac{1}{\mathrm{Re}} \frac{\partial}{\partial \Omega_1}\left(\langle\Delta \omega(\boldsymbol x,t)|\Omega_1, t\rangle f_1\right)-\frac{\partial}{\partial\Omega_1}\left(\left\langle\alpha(t)\Phi_1(\boldsymbol x,t)-\beta\Phi_2(\boldsymbol x,t)|\Omega_1,t\right\rangle f_1\right).
\end{equation}
with
\begin{equation}
    \Phi_1(\boldsymbol x,t)=\sum_{k_{f,\min} \leq |\boldsymbol{k}| \leq k_{f,\max}} 
    \hat{\omega}(\boldsymbol{k},t)\, e^{i \boldsymbol{k}\cdot\boldsymbol{x}}, \qquad \Phi_2(\boldsymbol x,t)=
    \sum_{k_{d,\min} < |\boldsymbol{k}| \leq k_{d,\max}} 
    \hat{\omega}(\boldsymbol{k},t)\, e^{i \boldsymbol{k}\cdot\boldsymbol{x}}.
\end{equation}

\section{Numerical method \& estimation of conditional averages} \label{sec:num}

The governing equations \eqref{eq:1pointPDF_HIT} and \eqref{eq:1pointPDF_HIT_forced} for the one-point vorticity PDF take the similar form of linear transport equations in sample space, with drift terms given by conditional expectations. The numerical solution therefore requires (\textit{i}) an accurate estimation of these conditional averages from DNS data, and (\textit{ii}) a stable and conservative discretization of the resulting transport equation in the vorticity variable.

\subsection{Estimation of conditional averages}

Equations \eqref{eq:1pointPDF_HIT} and \eqref{eq:1pointPDF_HIT_forced} are \textit{unclosed} with terms expressed as conditional expectations of the form
\begin{equation} \label{eq:con_averages}
    M_\nu := \langle \nu\Delta \omega(\boldsymbol x,t)|\Omega_1, t\rangle, \quad
    M_f: = \langle\alpha(t)\Phi_1(\boldsymbol x,t)|\Omega_1,t \rangle, \quad
    M_d: = \langle\beta\Phi_2(\boldsymbol x,t)|\Omega_1,t \rangle
\end{equation}
Here, we evaluate these quantities directly from DNS data using a kernel-based estimator of Nadaraya-Watson type, which is a prevalent method for estimating conditional expectations \cite{Bierens_1994}.
To do so, Given a set of samples $\{(\omega_i, g_i)\}_{i=1}^N$, where $g_i$ denotes the quantity of interest (e.g. $\Delta\Omega$, forcing contributions), the conditional expectation at a given value $\Omega_1$ is approximated by
\begin{equation} \label{eq:estimator}
\langle g \mid \omega(\boldsymbol x,t)=\Omega_1 \rangle
\approx
\frac{\sum_{i=1}^{N} K_h(\Omega_1 - \omega_i)\, g_i}
{\sum_{i=1}^{N} K_h(\Omega_1 - \omega_i)}.
\end{equation}
In the present work, we adopt the simplest possible kernel, namely an indicator (box) kernel,
\begin{equation}
K_h(\Omega_1 - \omega_i) =
\begin{cases}
1, & |\Omega_1 - \omega_i| \leq h/2, \\
0, & \text{otherwise},
\end{cases}
\end{equation}
which corresponds to a local binning procedure in $\omega$-space. This choice corresponds to a projection onto a finite-dimensional space of piecewise constant functions in $\omega$, ensures robustness and low computational cost, 
while avoiding additional smoothing parameters beyond the bin width $h$.

A key observation is that, in the case of statistical homogeneity, the DNS grid points at a fixed time snapshot can be treated as independent samples for constructing the estimator \eqref{eq:estimator}. The number of samples used in the estimator can be varied by subsampling the grid, allowing a systematic study of convergence with respect to sample size.

\subsection{Numerical solution of the governing equations}

To treat the governing equations \eqref{eq:1pointPDF_HIT} and \eqref{eq:1pointPDF_HIT_forced}, it is more convenient to reformulate the problems in terms of the cumulative distribution function (CDF)
\begin{equation}
F_1(\Omega_1,t) = \int_{-\infty}^{\Omega_1} f_1(\xi,t)\, d\xi,
\end{equation}
which satisfies a first-order transport equation of the form
\begin{equation}
\frac{\partial F_1}{\partial t} + M(\Omega_1,t)\,\frac{\partial F_1}{\partial \Omega_1} = 0,
\end{equation}
where the drift coefficient $M(\Omega_1,t):=M_\nu+M_f-M_d$ is given by the corresponding conditional averages, including viscous and forcing contributions from \eqref{eq:con_averages}.
Then, the transport equation for $F_1$ is solved using a characteristic-based approach, which is based on the elementary fact that the CDF is conserved along the characteristics
\begin{equation}
\frac{d\Omega_1}{dt} = M(\Omega_1(t),t).
\end{equation}
Therefore, after discretizing the computational domain using a uniform grid in $\Omega_1$-space into $\{\Omega_j\}$, we evaluate the solution at time $t^{n+1}$ as
\begin{equation}
F_1^{n+1}(\Omega_j) = F_1^n(\Omega_j^\ast),
\end{equation}
where $\Omega_j^\ast$, the foot of the characteristic that arrives at $\Omega_j$ at time $t^{n+1}$, is obtained by integrating the characteristic equation backward in time,
\begin{equation}
\Omega_j^\ast = \Omega_j - \Delta t\, M(\Omega_j,t^n).
\end{equation}
Linear interpolation is adopted to evaluate $F_1^n(\Omega_j^\ast)$.

For numerical implementation, we choose 200 (decaying HIT) or 400 (forced HIT) uniform grids to discretize $\Omega_1$. The lower and upper limits of the discretized $\Omega_j$'s are such chosen that $F_1 \to 0$ and $F_1 \to 1$ at the boundaries. The initial condition $F_1(\Omega_1,0)$ is obtained from DNS data by direct accumulation. At each time step, the drift term $M(\Omega_1,t^n)$ is interpolated for each $\Omega_j$ based on the re-constructed arrays. Finally, the PDF $f_1$ is recovered from the CDF $f_1 = \partial F_1/\partial \Omega_1$ using finite differences.

%===============================================
\section{Results and discussion} \label{sec:result}

\begin{figure}
    \centering
    \includegraphics[width=1\linewidth]{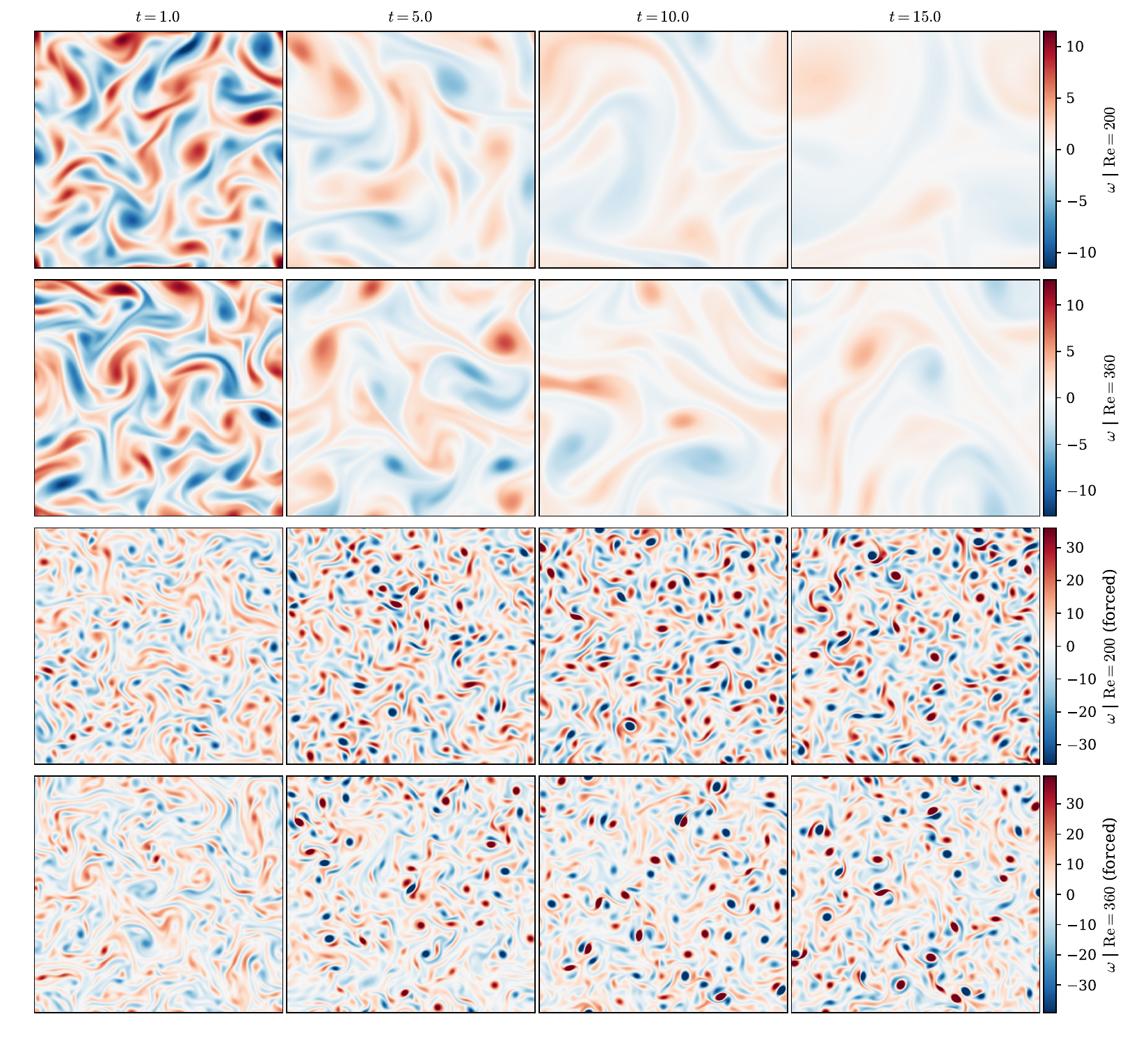}
    \caption{Snapshots of vorticity contours from the DNS results of the four cases listed in Table~\ref{tab:dns-params}.}
    \label{fig:dns}
\end{figure}

\subsection{Vorticity PDF for decaying 2D HIT}

\begin{figure}
\centering
\includegraphics[width=1.05\textwidth]{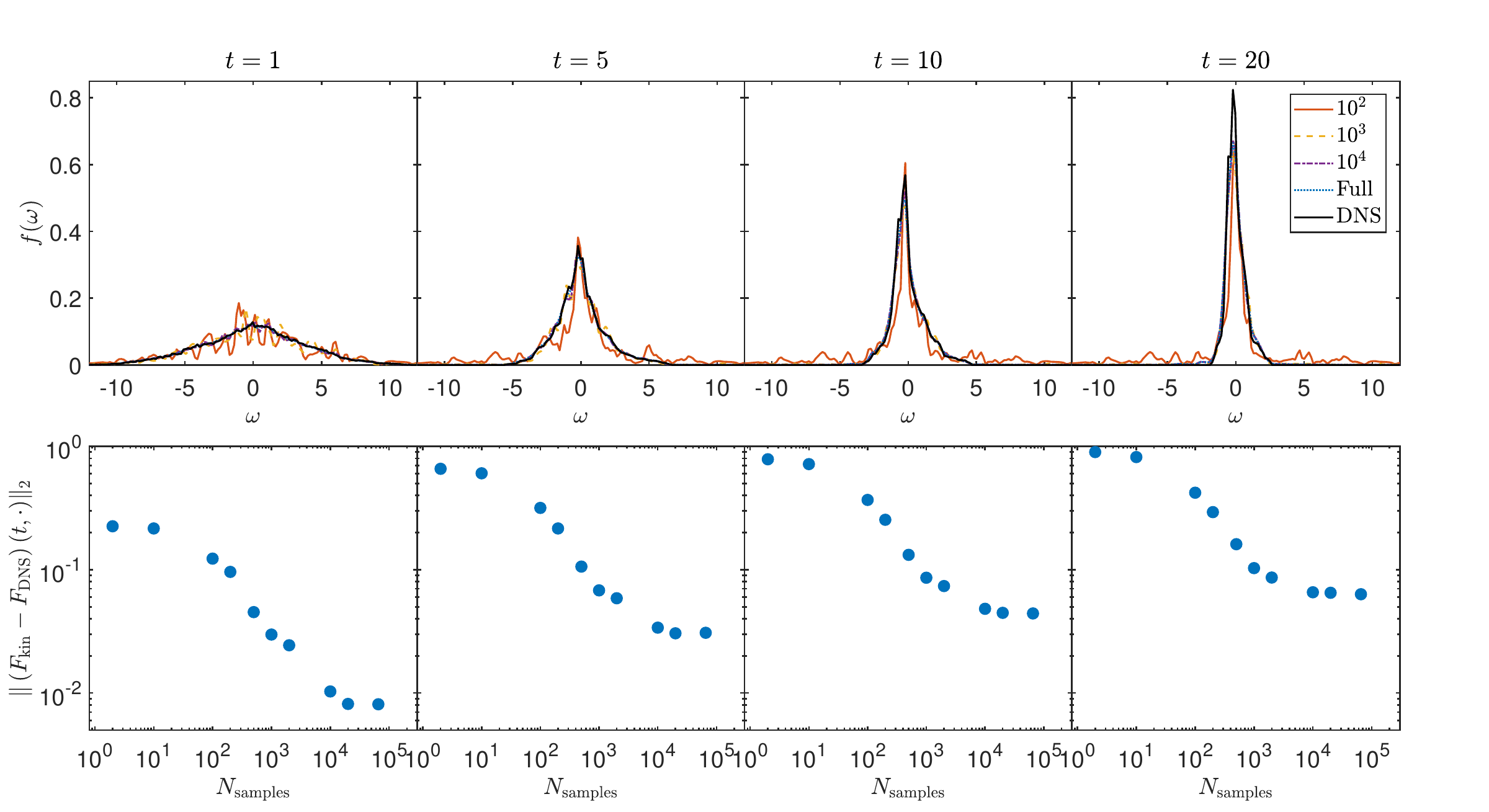}
\caption{The 2D decaying HIT with $\mathrm{Re}=200$. 
Upper: single-point vorticity PDFs $f(t,\omega)$ at different snapshots. Each plot presents the PDF reconstructed from DNS results of $\omega$ (using all grids, labelled as ``DNS") and those solved from the kinetic equation \eqref{eq:1pointPDF_HIT} using various numbers of samples (i.e., grids) to approximate the conditional expectation $M_\nu=\langle \nu\Delta \omega | \Omega_1,t \rangle$ from \eqref{eq:con_averages}. 
Lower: $L^2$-errors of the kinetic-derived CDFs (against the DNS-recovered ones) decrease with increasing numbers of samples for the conditional expectation at different times.} \label{fig:1point_decay_HIT_Re200}
\end{figure}

\begin{figure}
\centering
\includegraphics[width=1.05\textwidth]{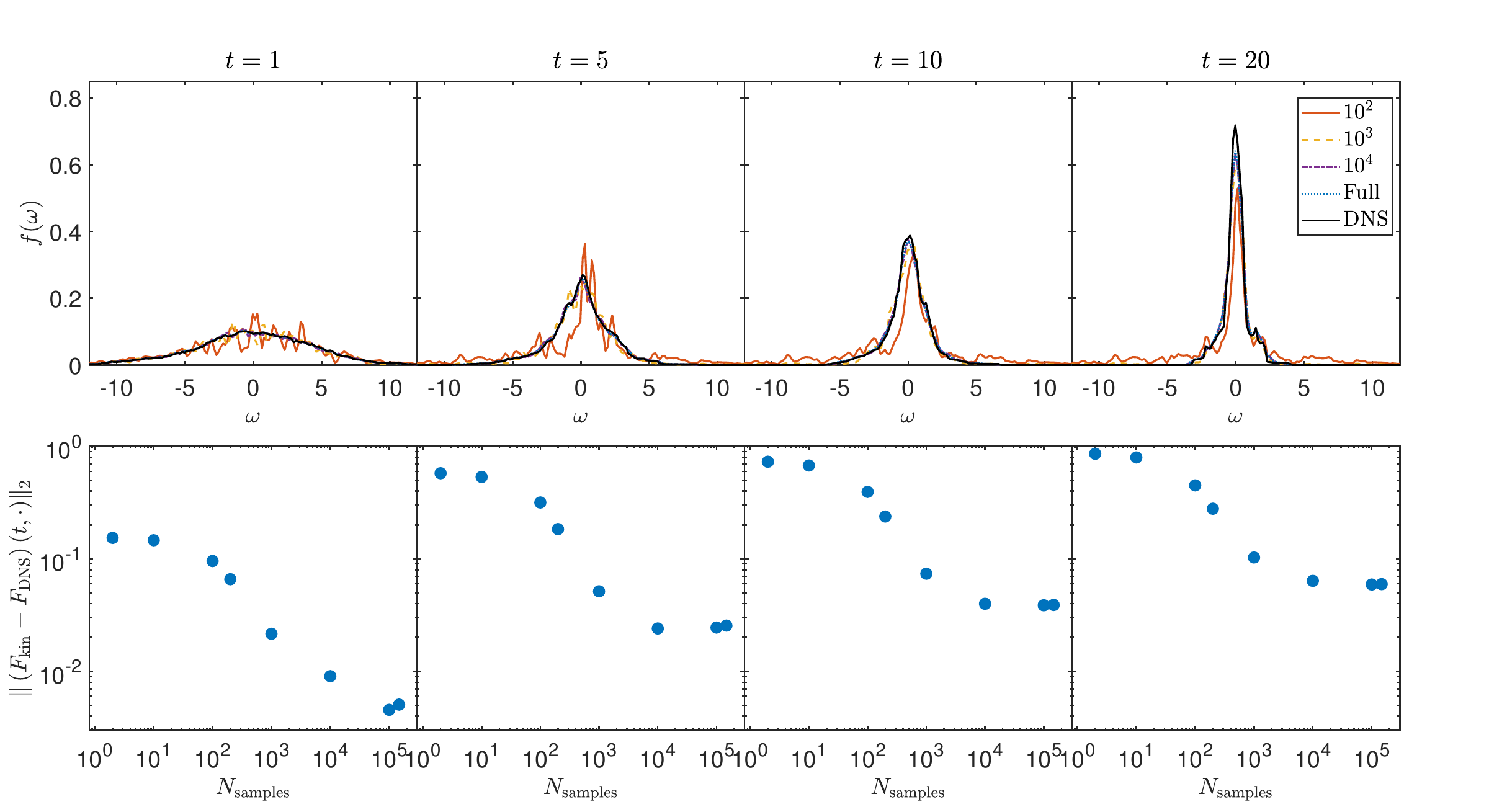}
\caption{The 2D decaying HIT with $\mathrm{Re}=360$. 
Upper: single-point vorticity PDFs $f(t,\omega)$ at different snapshots. Each plot presents the PDF reconstructed from DNS results of $\omega$ (using all grids, labelled as ``DNS") and those solved from the kinetic equation \eqref{eq:1pointPDF_HIT} using various numbers of samples (i.e., grids) to approximate the conditional expectation $M_\nu=\langle \nu\Delta \omega | \Omega_1,t \rangle$. 
Lower: $L^2$-errors of the kinetic-derived CDFs (against the DNS-recovered ones) reduce with increasing numbers of samples for the conditional expectation at different times.} \label{fig:1point_decay_HIT_Re360}
\end{figure}

\begin{figure}
\centering
\includegraphics[width=\textwidth]{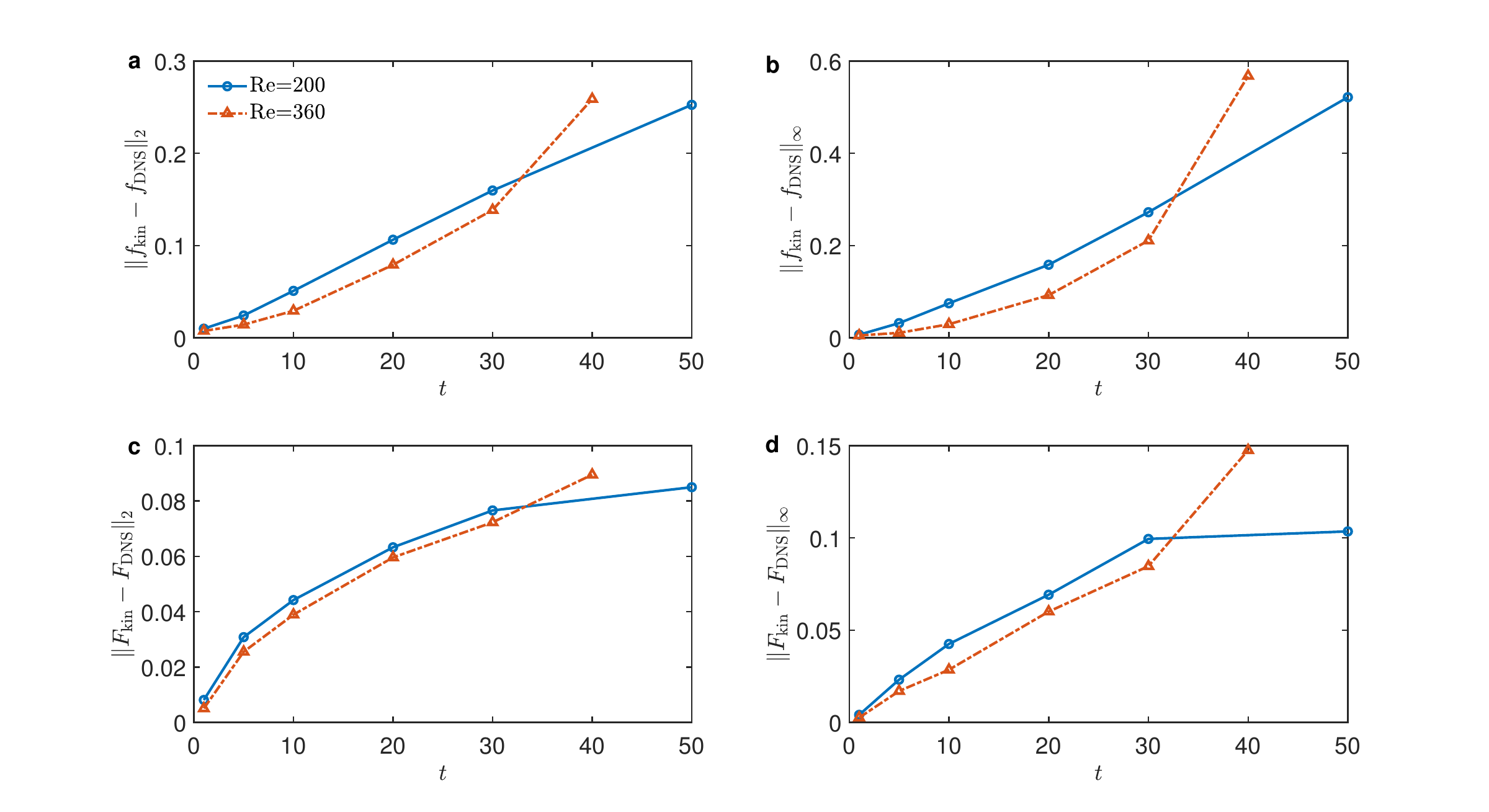}
\caption{Time evolution of the $L^2$- and $L^\infty$-errors of kinetic-derived PDFs $f(\Omega_1,t)$ and CDFs $F(\Omega_1,t)$ with respect to the distributions out of DNS statistics. Results are shown for 2D decaying HIT with $\mathrm{Re}=200$ and 360. All grids are used for the conditional averages.} \label{fig:errors_decay_HIT}
\end{figure}

We first consider the decaying homogeneous isotropic turbulence (HIT) cases without external forcing, 
governed by the PDF transport equation \eqref{eq:1pointPDF_HIT}. As seen in the first and second rows of Figure~\ref{fig:dns}, for both $Re=200$ and $Re=360$, the vorticity field `coarsens' and decays with time. The evolution of vorticity PDFs is shown in Figures~\ref{fig:1point_decay_HIT_Re200} and~\ref{fig:1point_decay_HIT_Re360}, 
respectively, while more quantitative metrics for the error are provided in Figure~\ref{fig:errors_decay_HIT}. 

The upper panels in Figures~\ref{fig:1point_decay_HIT_Re200} and~\ref{fig:1point_decay_HIT_Re360} display the time evolution of the single-point vorticity PDF $f_1(\Omega_1,t)$, comparing the distributions obtained directly from DNS with those reconstructed from the kinetic equation. At early times, the PDF is close to Gaussian due to the random initial condition. As the flow evolves, the PDF seems to deviate from Gaussianity. Indeed, the distributions become increasingly peaked around $\Omega_1=0$, accompanied by the development of heavier tails. This behavior could be related to the formation of coherent vortex structures, the 
associated intermittency in the vorticity field, and, most importantly, the decaying nature of the flow.
The kinetic approach reproduces this evolution with high fidelity. Even for moderate sample sizes used in the estimation of $M_\nu=\langle \nu\Delta \omega | \Omega_1,t \rangle$, the reconstructed PDFs capture both the central peak 
and the tail behavior accurately. Increasing the number of samples beyond $10^4$ leads to a systematic improvement, with the kinetic solution becoming nearly indistinguishable from the DNS reference.

The lower panels in Figures~\ref{fig:1point_decay_HIT_Re200} and~\ref{fig:1point_decay_HIT_Re360} quantify the accuracy of the method by showing the $L^2$-errors of the cumulative distribution functions (CDFs) derived from the kinetic solution, compared to those obtained directly from DNS. A clear convergence trend is observed: the error decreases monotonically as the number of samples used in the conditional average estimator increases. This implies that the dominant source of error arises from the statistical estimation of the conditional expectations.
We note that as $N_\mathrm{sample}>2\times 10^4$, the error becomes `saturated', which may be mainly attributed to a fixed value of the bin size $h$ used in the estimator \eqref{eq:estimator}. Moreover, the convergence behavior is consistent across different time instances, demonstrating that the estimator 
remains effective even as the PDF evolves and deviates significantly from Gaussianity.

The influence of the Reynolds number can be manifested through a comparison of the results for $Re=200$ and $Re=360$. Clearly, the overall behavior is qualitatively similar, while some quantitative 
differences can be observed. At higher Reynolds number, the PDFs exhibit slightly more pronounced tails, suggesting a slower decay 
of extreme vorticity events due to weaker viscous damping. Nevertheless, the kinetic method maintains its accuracy across both Reynolds numbers, demonstrating the approach to be robust against changes within the considered flow regimes.

It is already seen from Figures~\ref{fig:1point_decay_HIT_Re200} and~\ref{fig:1point_decay_HIT_Re360} that the errors generally grow with time. Figure~\ref{fig:errors_decay_HIT} provides a more detailed view, showing both $L^2$- and $L^\infty$-errors for the PDFs and CDFs, where the kinetic approach employs all grids to construct the estimator \eqref{eq:estimator}. The errors of $f_1$ exhibit substantial growth with time, consist with the observed trend that the kinetic-derived $f_1$ struggles to match the peak around $\Omega_1=0$. By contrast, the growth of $F_1$-errors are `sublinear' in time (Note that the $L^\infty$-error of $F_1$ must be bounded by 1). We remark that the error mainly comes from the inaccuracy in estimating the conditional averages.

\subsection{Vorticity PDF for forced 2D HIT}

\begin{figure}
\centering
\includegraphics[width=1.05\textwidth]{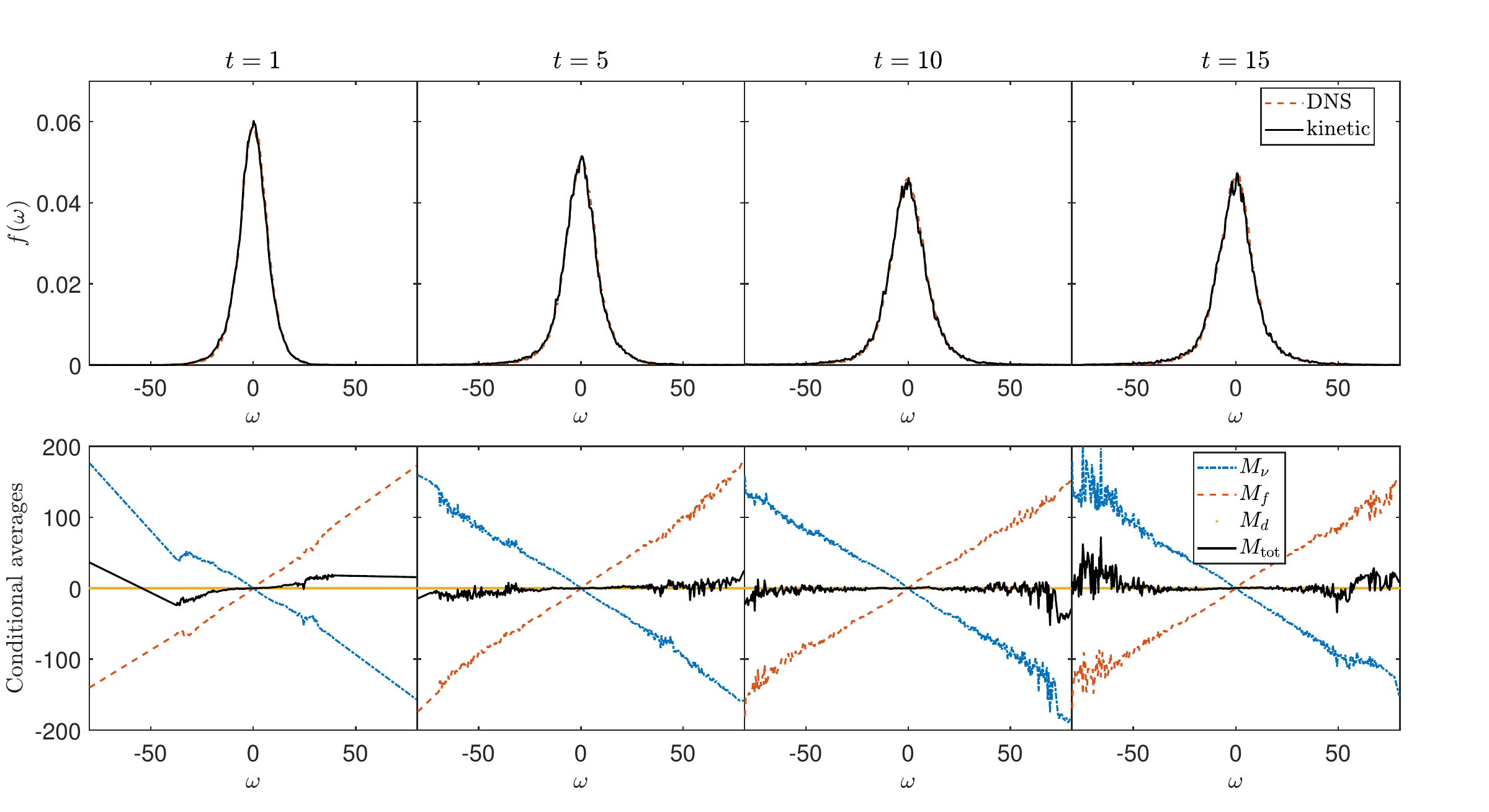}
\caption{The forced 2D HIT with $\mathrm{Re}=200$. 
Upper: single-point vorticity PDFs $f_1(\Omega_1,t)$ at different snapshots. 
Lower: The conditional averages at different times. All conditional averages are evaluated using the full set of grids.} \label{fig:forcedHIT_Re200}
\end{figure}

\begin{figure}
\centering
\includegraphics[width=1.05\textwidth]{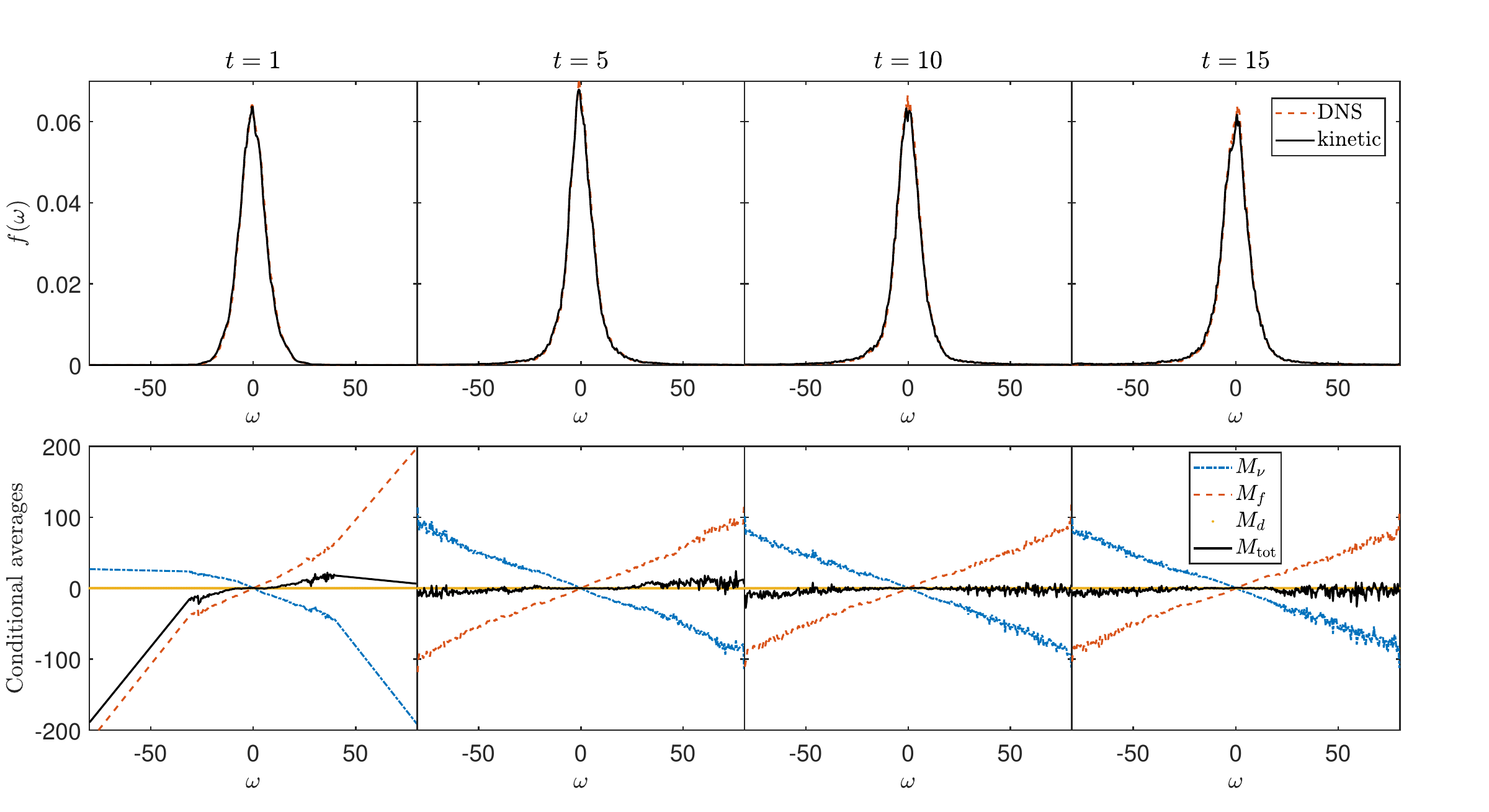}
\caption{The forced 2D HIT with $\mathrm{Re}=360$. 
Upper: single-point vorticity PDFs $f_1(\Omega_1,t)$ at different snapshots. 
Lower: The conditional averages at different times. All conditional averages are evaluated using the full set of grids.}
\label{fig:forcedHIT_Re360}
\end{figure}

We now consider the forced HIT cases, governed by the extended PDF transport equation \eqref{eq:1pointPDF_HIT_forced}, 
in which external forcing and large-scale damping balance viscous dissipation. See the detailed descriptions of energy injection in Section~\ref{sec:DNS}. As seen in the third and fourth rows of Figure~\ref{fig:dns}, for both $Re=200$ and $Re=360$, the vorticity field does not decay, with a typical length scale endures. The corresponding evolution of vorticity PDFs is shown in Figures~\ref{fig:forcedHIT_Re200} and~\ref{fig:forcedHIT_Re360}.

The upper panels in Figures~\ref{fig:forcedHIT_Re200} and~\ref{fig:forcedHIT_Re360} display the time evolution of the single-point vorticity PDF $f_1(\Omega_1,t)$. In contrast to the decaying case, the distributions gradually approach a statistically stationary state, with only minor fluctuations at later times. The resultant stationary PDFs are considerably `wider' than those from the decaying cases, indicating continuously-populated extreme vorticity events (with, say, $|\omega|>50$) due to a sustained injection of fluctuations by the forcing mechanism. Clearly, the kinetic solutions obtained from \eqref{eq:1pointPDF_HIT_forced} accurately reproduce both the transient evolution and the stationary  distributions, showing excellent agreement with the DNS results.

In such forcing cases, it is illustrative to look closer into the conditional averages entering the drift term in \eqref{eq:1pointPDF_HIT_forced}, as shown in the lower panels in Figures~\ref{fig:forcedHIT_Re200} and~\ref{fig:forcedHIT_Re360}. It turns out that the viscous contribution $M_\nu:=\langle \nu\Delta \omega \mid \omega=\Omega_1 \rangle$ exhibits an approximately linear dependence on $\omega$ with negative slopes, corresponding to an effective dissipative drift towards $\Omega_1=0$. The greater fluctuations for large $|\omega|>50$ is a direct consequence of statistical noise due to limited samples. The \textit{ad hoc} forcing contribution $M_f$ in \eqref{eq:con_averages} provides a compensating effect, injecting fluctuations that effectively counterbalance viscous 
dissipation. The dissipation term $M_d$ plays a negligible role as it is several orders smaller than $M_\nu$ and $M_f$. In the statistically stationary regime, the combined drift term leads to an approximate balance in $\omega$-space (namely, $M_\mathrm{total}\approx 0$), which is directly reflected in the time-invariant shape of the PDF. 
Therefore, the forced HIT results highlight the capability of the present framework to incorporate complex dynamical effects through data-driven conditional averages.

\section{Conclusion} \label{sec:conclusion}

In this work, we have developed a hybrid numerical framework for the solution of one-point vorticity PDF transport equations in two-dimensional homogeneous isotropic turbulence, for which the governing equations reduce to linear transport equations in sample space with unclosed conditional averages. These terms are then approximated directly from DNS data using a non-parametric Nadaraya-Watson-type estimator, while the resulting transport equations are solved in terms of the cumulative distribution function using a characteristic-based method.

The numerical results for both decaying and forced turbulence demonstrate that the proposed approach accurately captures the time evolution of the vorticity PDF. In the decaying case, the method reproduces the transition from near-Gaussian initial conditions to distributions more concentrated at zero. In the forced case, it correctly recovers statistically stationary PDFs and provides direct insight into the dynamical balance between viscous dissipation and external forcing through the structure of the conditional averages. A key observation is that the overall accuracy is primarily controlled by the estimation of the conditional averages. 

This work provides a natural bridge between 
data-driven closure strategies and the theory of turbulent flows. While the present study focuses on one-point statistics, the approach readily extends to multi-point PDFs, which will be addressed in future work.

\FloatBarrier
\section*{Acknowledgments} 
Financial support by the German Research Foundation (DFG), within  the project No. 526024901 of the Priority Programme - SPP 2410 Hyperbolic Balance Laws in Fluid Mechanics: Complexity, Scales, Randomness (CoScaRa) is acknowledged.

\bibliography{wileyNJD-AMS.bib}

\end{document}